# Heuristic Contraction Hierarchies with Approximation Guarantee

Robert Geisberger

Karlsruhe Institute of Technology (KIT), 76128 Karlsruhe, Germany, `geisberger@kit.edu`

August 21, 2018

**Abstract**

We present a new heuristic point-to-point routing algorithm based on contraction hierarchies (CH). Given an $\varepsilon \geq 0$, we can prove that the length of the path computed by our algorithm is at most $(1 + \varepsilon)$ times the length of the optimal (shortest) path. CH is based on *node contraction*: removing nodes from a network and adding shortcut edges to preserve shortest path distances. Our algorithm tries to avoid shortcuts even when a replacement path is $\varepsilon$ times longer.

## 1 Introduction

Routing in static road networks is essentially solved. There exist fast routing algorithms that are exact. However, for other graph classes, e.g. unit disk graphs as they appear in sensor networks, the algorithms do not work very well. Also when several objective functions should be supported within a road network, current algorithms [4] face some problems since the hierarchy of the network is different for different objective functions, e.g. time and distance. One possibility to alleviate these problems is to drop the exactness of the algorithms and allow some error. We show how to adapt contraction hierarchies (CH) [5] so that we can guarantee an error of $\varepsilon$.

#### Related Work

There has been extensive work on speed-up techniques for road networks. We refer to [2, 6, 1] for an overview.

We can classify current algorithms into three categories: *hierarchical* algorithms, *goal-directed* approaches and combinations of both. Our algorithm is based on contraction hierarchies (CH) [5]. A CH uses a single node order and contracts the nodes in this order. A slightly modified bidirectional Dijkstra shortest path search then answers a query request, touching only a few hundred nodes. For our algorithm, we modify the node contraction, i.e. the decision which shortcuts we have to add, and the query.

Recently, also speed-up techniques for *time-dependent* road networks have been developed ([3] presents an overview).

Also algorithms to support *multiple* objective function have been developed lately. The most recent one is an adaption of the CH algorithm [4] to allow a flexible query.

## 2 Heuristic Node Contraction

CH performs precomputation on a directed graph $G = (V, E)$ with edge weight function $c : E \to \mathbb{R}_+$. The precomputation of CH is based on *node contraction*: a node $u$ is contracted by removing it from the network in such a way that shortest paths in the remaining graph are preserved. In the exact scenario, where we want to find the optimal path, we use Algorithm 1. To preserve all shortest path distances, it is enough to preserve the shortest paths distances between the neighbors of $u$. So given two neighbors $v$ and $w$ with edges $(v, u)$ and $(u, w)$, we should find the shortest path $P$ between $v$ and $w$ avoiding $u$ (Line 4). When the length of $P$ is longer than the length of the path $\langle v, u, w \rangle$, a shortcut between $v$ and $w$ is necessary with weight $c(v, u) + c(u, w)$. When the length of $P$ is not longer, no shortcut is necessary and $P$ witnesses this, in this case we call $P$ a *witness*.

In the heuristic scenario, we use Algorithm 2. Intuitively, we want to avoid more shortcuts than in the exact scenario, i.e. a shortcut between $v$ and $w$ should be avoided, even when the path $P$ is a bit longer than $\langle v, u, w \rangle$. To be able to guarantee a maximum relative error of $\varepsilon$, we need to ensure that the errors do not stack when a node on $P$ is contracted later. So we store another edge weight $\tilde{c}$ with each edge. Intuitively, $\tilde{c}$ is the witness memory of an edge. When a witness $P$ prevents a shortcut, even though in the exact scenario, the shortcut would be necessary, the witness must remember this *delta*. We let the edges of the witness remember this (Lines 9–11), so that $\tilde{c}(P) \leq \tilde{c}(v, u) + \tilde{c}(u, w)$. Note that $\gamma \leq \varepsilon$ (Line 9), which gives Lemma 1.

**Lemma 1** *For each edge $(v, w)$ holds*

$$\frac{c(v, w)}{1 + \varepsilon} \leq \tilde{c}(v, w) \leq c(v, w) \ .$$

Note that there are other ways to ensure that the errors do not stack. Currently, we proportionally distribute the delta among all edges of the witness. We could distribute the delta differently or even try to find other potential witnesses. Also, avoiding a shortcut can lead to more shortcuts later, as every shortcut is a potential witness later.

---
**Algorithm 1:** SimplifiedExactConstructionProcedure($G = (V, E)$,<)
---
**1 foreach** $u \in V$ *ordered by* $<$ *ascending* **do**
**2**     **foreach** $(v, u) \in E$ *with* $v > u$ **do**
**3**        **foreach** $(u, w) \in E$ *with* $w > u$ **do**
**4**           find shortest path $P = \langle v, \ldots, w \rangle \not\ni u$;
**5**           **if** $c(P) > c(v, u) + c(u, w)$ **then**
**6**              $E := E \cup \{(v, w)\}$ (use weight $c(v, w) := c(v, u) + c(u, w)$)

---
**Algorithm 2:** SimplifiedHeuristicConstructionProcedure($G = (V, E)$,<,$\varepsilon$)
---
**1** $\tilde{c} := c$;          // store second weight per edge
**2 foreach** $u \in V$ *ordered by* $<$ *ascending* **do**
**3**     **foreach** $(v, u) \in E$ *with* $v > u$ **do**
**4**        **foreach** $(u, w) \in E$ *with* $w > u$ **do**
**5**           find shortest path $P = \langle v, \ldots, w \rangle \not\ni u$;
**6**           **if** $c(P) > (1 + \varepsilon)(\tilde{c}(v, u) + \tilde{c}(u, w))$ **then**
**7**              $E := E \cup \{(v, w)\}$ (use weight $c(v, w) := c(v, u) + c(u, w)$, $\tilde{c}(v, w) := \tilde{c}(v, u) + \tilde{c}(u, w)$);
**8**           **else**
**9**              $\gamma := \frac{c(P)}{\tilde{c}(v,u) + \tilde{c}(u,w)} - 1$;          // $c(P) = (1 + \gamma)(\tilde{c}(v, u) + \tilde{c}(u, w))$
**10**              **foreach** $(x, y) \in P$ **do**
**11**                 $\tilde{c}(x, y) := \min\left\{\tilde{c}(x, y), \frac{c(x,y)}{1+\gamma}\right\}$;

## 3 Heuristic Query

The basic query algorithm is the same as for exact CH (Algorithm 3). It is a symmetric Dijkstra-like bidirectional procedure. We will introduce it formally and prove its correctness.

The query algorithm does not relax edges leading to nodes lower than the current node. This property is reflected in the *upward graph* $G_\uparrow := (V, E_\uparrow)$ with $E_\uparrow := \{(u, v) \in E \mid u < v\}$ and, analogously, the *downward graph* $G_\downarrow := (V, E_\downarrow)$ with $E_\downarrow := \{(u, v) \in E \mid u > v\}$).

We perform a forward search in $G_\uparrow$ and a backward search in $G_\downarrow$. Forward and backward search are interleaved, we keep track of a tentative shortest-path length and abort the forward/backward search process when all keys in the respective priority queue are greater than the tentative shortest-path length (abort-on-success criterion).

Both search graphs $G_\uparrow$ and $G_\downarrow$ can be represented in a single, space-efficient data structure: an adjacency array. Each node has its own edge group of incident edges. Since we perform a forward search in $G_\uparrow$ and a backward search in $G_\downarrow$, we only need to store an edge in the edge group of the lower incident node. This formally results in a search graph $G^* = (V, E^*)$ with $\overline{E_\downarrow} := \{(v,u) \mid (u,v) \in E_\downarrow\}$ and $E^* := E_\uparrow \cup \overline{E_\downarrow}$. Finally, we introduce a forward and a backward flag such that for any edge $e \in E^*$, $\uparrow(e) = $ true iff $e \in E_\uparrow$ and $\downarrow(e) = $ true iff $e \in \overline{E_\downarrow}$. Note that $G^*$ is a directed acyclic graph.

**Algorithm 3:** ExactQuery($s,t$)

1 $d_\uparrow := \langle \infty, \ldots, \infty \rangle$; $d_\uparrow[s] := 0$; $d_\downarrow := \langle \infty, \ldots, \infty \rangle$; $d_\downarrow[t] := 0$, $d := \infty$;  // tentative distances
2 $Q_\uparrow = \{(0,s)\}$; $Q_\downarrow = \{(0,t)\}$; $r := \uparrow$;  // priority queues
3 **while** ($Q_\uparrow \neq \emptyset$ or $Q_\downarrow \neq \emptyset$) **and** ($d > \min\{\min Q_\uparrow, \min Q_\downarrow\}$) **do**
4     **if** $Q_{\neg r} \neq \emptyset$ **then** $r := \neg r$;  // interleave direction, $\neg \uparrow = \downarrow$ and $\neg \downarrow = \uparrow$
5     $(\cdot, u) := Q_r.\text{deleteMin}()$; $d := \min\{d, d_\uparrow[u] + d_\downarrow[u]\}$;  // $u$ is settled and new candidate
6     **foreach** $e = (u,v) \in E^*$ **do**  // relax edges of $u$
7        **if** $r(e)$ **and** $(d_r[u] + c(e) < d_r[v])$ **then**  // shorter path found
8           $d_r[v] := d_r[u] + c(e)$;  // update tentative distance
9           $Q_r.\text{update}(d_r[v],v)$;  // update priority queue

10 **return** $d$;

**Lemma 2** *Let $P$ be an s-t-path $P$ in the heuristic CH $(G = (V,E), <, \varepsilon)$. Then there exists an s-t-path $P'$ in the CH of form $\langle s = u_0, u_1, \ldots, u_p, \ldots, u_q = t\rangle$ with $p, q \in \mathbb{N}$, $u_i < u_{i+1}$ for $i \in \mathbb{N}, i < p$ and $u_j > u_{j+1}$ for $j \in \mathbb{N}, p \leq j < q$, short **p**ath **f**orm (PF) And for $P'$ holds $\tilde{c}(P') \leq \tilde{c}(P)$.*

*Proof.* Given a shortest $s$-$t$-path $P = \langle s = u_0, u_1, \ldots, u_p, \ldots, u_q = t \rangle$ with $p, q \in \mathbb{N}$, $u_p = \max P$, that is not of the form (PF). Then there exists a $k \in \mathbb{N}, k < q$ with $u_{k-1} > u_k < u_{k+1}$. We will recursively construct a path of the form (PF).

Let $M_P := \{u_k \mid u_{k-1} > u_k < u_{k+1}\}$ denote the set of local minima excluding nodes $s, t$. We show that there exists a $s$-$t$-path $P'$ with $M_{P'} = \emptyset$ or $\min M_{P'} > \min M_P$.

Let $u_k := \min M_P$ and consider the two edges $(u_{k-1}, u_k), (u_k, u_{k+1}) \in E$. Both edges already exist at the beginning of the contraction of node $u_k$. So there is either a witness path $Q = \langle u_{k-1}, \ldots, u_{k+1} \rangle$ consisting of nodes higher than $u_k$ with $c(Q) \leq (1+\varepsilon)(\tilde{c}(u_{k-1}, u_k) + \tilde{c}(u_k, u_{k+1}))$ or a shortcut $(u_{k-1}, u_{k+1})$ of the same weight is added. So the subpath $P|_{u_{k-1} \to u_{k+1}}$ can either be replaced by $Q$ or by the shortcut $(u_{k-1}, u_{k+1})$. If we replace the subpath by $Q$, our construction ensures that $\tilde{c}(Q) \leq \tilde{c}(P|_{u_{k-1} \to u_{k+1}})$. Also if we added a shortcut, $\tilde{c}(u_{k-1}, u_{k+1}) \leq \tilde{c}(P|_{u_{k-1} \to u_{k+1}})$ holds. So the resulting path $P'$ consists of nodes higher than $u_k$ and has the property $\tilde{c}(P') \leq \tilde{c}(P)$. Since $n < \infty$, there must exist a $s$-$t$-path $P''$ with $M_{P''} = \emptyset$, $\tilde{c}(P'') \leq \tilde{c}(P)$ and of the form described in (PF). □

Theorem 1 proofs the correctness of our basic query algorithm.

**Theorem 1** *Given a source node $s$ and a target node $t$. Let $\tilde{d}(s,t)$ be the distance computed by the heuristic CH algorithm with $\varepsilon \geq 0$ and let $d(s,t)$ be the optimal (shortest) distance in the original graph. Then $d(s,t) \leq \tilde{d}(s,t) \leq (1+\varepsilon)d(s,t)$.*

*Proof.* Let $(G = (V,E), <, \varepsilon)$ be a heuristic CH with $\varepsilon \geq 0$. Let $s, t \in V$ be the *source/target* pair of a query. It follows from the definition of a shortcut, that the shortest path distance between $s$ and $t$ in the CH is the same as in the original graph. So $d(s,t) \leq \tilde{d}(s,t)$ holds. Every shortest $s$-$t$-path in the original graph still exists in the CH but there may be additional $s$-$t$-paths. However since we use a modified Dijkstra algorithm that does not relax all incident edges of a settled node, our query algorithm does only find particular ones. In detail, exactly the shortest paths of the form (PF) are found by our query algorithm. From Lemma 2, we know that if there exists a shortest $s$-$t$-path $P$ then there also exists an $s$-$t$-path $P'$ of the form (PF) with $\tilde{c}(P') \leq \tilde{c}(P)$. Because of Lemma 1, we know that $\frac{c(P')}{1+\varepsilon} \leq \tilde{c}(P')$ and $\tilde{c}(P) \leq c(P)$ so that $c(P') \leq (1+\varepsilon)c(P)$. So our query algorithm will either find $P'$ or another path, that is not longer than $P'$. □

Although we use $\tilde{c}$ in the proof, the query algorithm does not use it at all. So we only need to store it during precomputation but we do not need to store it for the query.

## 4 Heuristic Stall-on-Demand

In the previous section, we proved that the basic heuristic query algorithm does not need any changes compared to the exact scenario. However, an important ingredient of a practically efficient query algorithm is the stall-on-demand technique first described by Schultes et al. [7]. This single improvement brings additional speedup of factor two or more.

While performing a query, we do not relax all edges of a node, we only relax edges leading upwards. Here and in the following we well assume that the query is from source node $s$ to target node $t$ and we will only consider the forward search. The backward search is completely symmetric to the forward search. We call a path leading only upwards being an *upward* path. In the exact scenario, we check, whether a settled node $u$ in the forward search is reached suboptimally via a $s$-$u$-path $P = \langle s = v_1, \ldots, v_k = u \rangle$, i.e if there is a $s$-$u$-path $P'$ with $c(P) > c(P')$. If this was the case, we can safely stop (*stall*) the search at $u$ since it would never contribute to a shortest path. The path $P$ is an upward path and stalling happens if and only if the path $P'$ is not an upward path..

However, in the heuristic scenario with $\varepsilon > 0$, we would destroy the correctness of our algorithm when we would apply the same rule. Consider as example the graph in Figure 1. During the contraction of $u$, no shortcut for the path $\langle x, u, v \rangle$ is added since the path $\langle x, y, v \rangle$ is a witness that is just a factor $1 + \varepsilon$ larger. The forward query starting at $s$ should settle the nodes in the order $s, x, y, u, v, z$. However, if we would not change the stalling condition, we would stall $u$ while settling it because the path $\langle s, u \rangle$ is longer than the path $\langle s, x, u \rangle$, which is not an upward path. Furthermore, we would propagate the stalling information to $v$, so node $v$ reached via upward path $P = \langle s, x, y, v \rangle$ gets stalled by the shorter path $P' = \langle s, x, u, v \rangle$. So node $v$ is stalled and we would never reach node $z$ with the forward search and therefore could never meet with the backward search there.

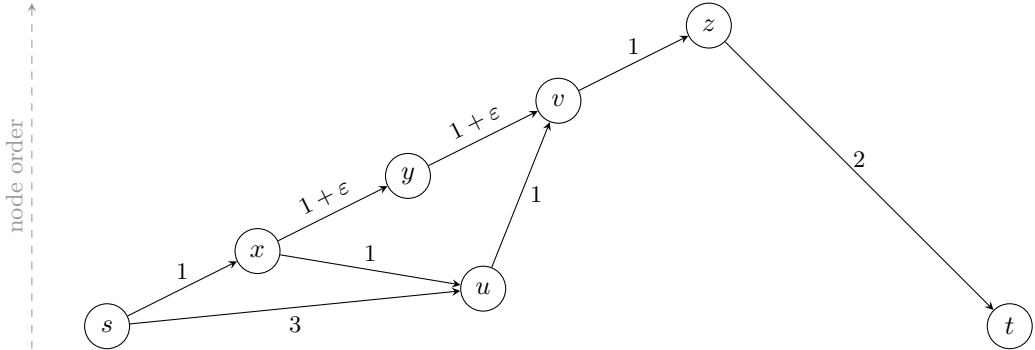

Figure 1: The stalling condition of the exact query fails in a heuristic CH, as node $z$ is never reached in the forward search from $s$.

To ensure the correctness, we change the stalling condition. We split the path $s$-$u$-path $P'$ in a path $P'_1$ and $P'_2$ so that $P'_1$ is the maximal upward subpath. Let $x$ be the node that splits $P'$ in these two parts. Then we stall $u$ only if node $x$ is reached by the forward search and

$$c(P'_1) + (1+\varepsilon)c(P'_2) < c(P) \tag{1}$$

The symmetric condition applies to the backward search.

**Algorithm 4:** ExactQuery(s,t)

1. $d_\uparrow := \langle \infty, \ldots, \infty \rangle$; $d_\uparrow[s] := 0$; $d_\downarrow := \langle \infty, \ldots, \infty \rangle$; $d_\downarrow[t] := 0$, $d := \infty$;    // tentative distances
2. $Q_\uparrow = \{(0, s)\}$; $Q_\downarrow = \{(0, t)\}$; $r := \uparrow$;    // priority queues
3. **while** $(Q_\uparrow \neq \emptyset$ **or** $Q_\downarrow \neq \emptyset)$ **and** $(d > \min\{\min Q_\uparrow, \min Q_\downarrow\})$ **do**
4.    **if** $Q_{\neg r} \neq \emptyset$ **then** $r := \neg r$;    // interleave direction, $\neg \uparrow = \downarrow$ and $\neg \downarrow = \uparrow$
5.    $(\cdot, u) := Q_r.\text{deleteMin}()$; $d := \min\{d, d_\uparrow[u] + d_\downarrow[u]\}$;    // $u$ is settled and new candidate
6.    **if** $isStalled(r, u)$ **then continue**;    // do not relax edges of a stalled node
7.    **foreach** $e = (u, v) \in E^*$ **do**    // relax edges of $u$
8.      **if** $r(e)$ **and** $(d_r[u] + c(e) < d_r[v])$ **then**    // shorter path found
9.         $d_r[v] := d_r[u] + c(e)$;    // update tentative distance
10.         $Q_r.\text{update}(d_r[v], v)$;    // update priority queue
11.         **if** $isStalled(r, v)$ **then** $unstall(r, v)$;
12.      **if** $(\neg r)(e)$ **and** $(d_r[v] + c(e) < d_r[u])$ **then**    // path via $v$ is shorter
13.         $stall(r, u, d_r[v] + c(e))$;    // stall $u$ with stalling distance $d_r[v] + c(e)$
14.         **break**;    // stop relaxing edges of stalled node $u$
15. **return** $d$;

**Algorithm 5:** HeuristicQuery(s,t)

1. $d_\uparrow := \langle \infty, \ldots, \infty \rangle$; $d_\uparrow[s] := 0$; $d_\downarrow := \langle \infty, \ldots, \infty \rangle$; $d_\downarrow[t] := 0$, $d := \infty$;    // tentative distances
2. $Q_\uparrow = \{(0, s)\}$; $Q_\downarrow = \{(0, t)\}$; $r := \uparrow$;    // priority queues
3. **while** $(Q_\uparrow \neq \emptyset$ **or** $Q_\downarrow \neq \emptyset)$ **and** $(d > \min\{\min Q_\uparrow, \min Q_\downarrow\})$ **do**
4.    **if** $Q_{\neg r} \neq \emptyset$ **then** $r := \neg r$;    // interleave direction, $\neg \uparrow = \downarrow$ and $\neg \downarrow = \uparrow$
5.    $(\cdot, u) := Q_r.\text{deleteMin}()$; $d := \min\{d, d_\uparrow[u] + d_\downarrow[u]\}$;    // $u$ is settled and new candidate
6.    **if** $isStalled(r, u)$ **then continue**;    // do not relax edges of a stalled node
7.    **foreach** $e = (u, v) \in E^*$ **do**    // relax edges of $u$
8.      **if** $r(e)$ **and** $(d_r[u] + c(e) < d_r[v])$ **then**    // shorter path found
9.         $d_r[v] := d_r[u] + c(e)$;    // update tentative distance
10.         $Q_r.\text{update}(d_r[v], v)$;    // update priority queue
11.         **if** $isStalled(r, v)$ **then** $unstall(r, v)$;
12.      **if** $(\neg r)(e) \wedge d_r[v] + (1+\varepsilon)c(e) < d_r[u]$ **then**    // path via $v$ is shorter
13.         $stall(r, u, d_r[v] + (1+\varepsilon)c(e))$;    // stall $u$ with stalling distance $d_r[v] + (1+\varepsilon)c(e)$
14.         **break**;    // stop relaxing edges of stalled node $u$
15. **return** $d$;

To prove that the stall-on-demand with (1) is correct, we will iteratively construct in Lemma 3 a new path from a stalled one.

**Lemma 3** *Let $(P, v, w)$ be a **s**tall **s**tate **t**riple (SST): $P$ being an s-t-path of the form (PF), node $v$ being reached by the forward search by $P|_{s \to v}$ and not stalled and node $w$ being reached by the backward search by $P|_{w \to t}$ and not stalled. Define a function $g$ on an SST:*

$$g(P, v, w) := c(P|_{s \to v}) + (1 + \varepsilon)\tilde{c}(P|_{v \to w}) + c(P|_{w \to t}).$$

*If one of the nodes in $P|_{v \to w}$ becomes stalled, then there exists an SST $(Q, x, y)$ with*

$$g(Q, x, y) < g(P, v, w).$$

*Proof.* Let $u \in P|_{v \to w}$ be the node that becomes stalled. WLOG we assume that $P|_{v \to u}$ is an upward path, i.e. the stalling happens during the forward search. Then there exists an $s$-$u$-path $P'$ that is split in $P'_1$ and $P'_2$ as defined in (1) so that $c(P'_1) + (1+\varepsilon)c(P'_2) < c(P|_{s \to u})$. Let $x$ be the node that splits $P'$ into these two subpaths. Let $R$ be the path of form (PF) that is constructed following Lemma 2 from the concatenation of $P'_2$ and $P|_{u \to w}$. Let $Q$ be the concatenation of $P'_1$, $R$ and $P|_{w \to t}$ and $y := w$. By construction, $(Q, x, y)$ is a SST and we will prove that it is the one that we are looking for.

$$\begin{aligned}
g(Q, x, y) &= c(Q|_{s \to x}) + (1+\varepsilon)\tilde{c}(Q|_{x \to y}) + c(Q|_{y \to t}) \\
&\stackrel{\text{def.}}{=} c(P'_1) + (1+\varepsilon)\tilde{c}(R) + c(P|_{w \to t}) \\
&\stackrel{\text{L.2}}{\leq} c(P'_1) + (1+\varepsilon)(\tilde{c}(P'_2) + \tilde{c}(P|_{u \to w})) + c(P|_{w \to t}) \\
&\stackrel{\text{L.1}}{\leq} c(P'_1) + (1+\varepsilon)c(P'_2) + (1+\varepsilon)\tilde{c}(P|_{u \to w}) + c(P|_{w \to t}) \\
&\stackrel{(1)}{<} c(P|_{s \to u}) + (1+\varepsilon)\tilde{c}(P|_{u \to w}) + c(P|_{w \to t}) \\
&= c(P|_{s \to v}) + c(P|_{v \to u}) + (1+\varepsilon)\tilde{c}(P|_{u \to w}) + c(P|_{w \to t}) \\
&\stackrel{\text{L.1}}{\leq} c(P|_{s \to v}) + (1+\varepsilon)\tilde{c}(P|_{v \to u}) + (1+\varepsilon)\tilde{c}(P|_{u \to w}) + c(P|_{w \to t}) \\
&= g(P, v, w)
\end{aligned}$$

□

With Lemma 3 we are able to prove the correctness of heuristic stall-on-demand (1) in Theorem 2.

**Theorem 2** *Theorem 1 still holds when we use heuristic stall-on-demand (1).*

*Proof.* The proof will iteratively construct SST's with Lemma 3 starting with the path $P$ found in the proof of Theorem 1/Lemma 2 and the nodes $s$ and $t$. Obviously, at the beginning of the query, both nodes $s$ and $t$ are reached and not stalled, so $(P, s, t)$ is an SST and

$$g(P, s, t) = c(P|_{s \to s}) + (1+\varepsilon)\tilde{c}(P|_{s \to t}) + c(P|_{t \to t}) = (1+\varepsilon)\tilde{c}(P) \leq (1+\varepsilon)d(s,t).$$

We will prove that after a finite number of applications of Lemma 3, we obtain an SST $(Q, x, y)$ so that $Q$ is found by our query with stalling. For this path $Q$ holds:

$$\begin{aligned}
c(Q) &= c(Q|_{s \to x}) + c(Q|_{x \to y}) + c(Q|_{y \to t}) \\
&\stackrel{\text{L.1}}{\leq} c(Q|_{s \to x}) + (1+\varepsilon)\tilde{c}(Q|_{x \to y}) + c(Q|_{y \to t}) \\
&= g(Q, x, y) \\
&\stackrel{\text{L.3}}{\leq} g(P, s, t) \\
&\leq (1+\varepsilon)d(s,t)
\end{aligned}$$

Since our graph is finite, and due to the "<" in Lemma 3, we can apply Lemma 3 only finitely many times. The final SST $(Q, x, y)$ will be found by our query algorithm since $x$ is is reached in the forward search and not stalled, $y$ is reached in the backward search and not stalled. And since this is the final SST, no node on the path $Q|_{x \to y}$ will be stalled so our query will find the path $Q$ or a shorter path. □

## 5 Applications

We described our heuristic for just a single edge weight function. Further applications may be time-dependent road networks, where not only the travel time functions are approximations, but also the shortcuts. Also, it can help for multi-criteria optimization. It would be simple to extend it to the flexible scenario [4] with two edge weight functions. This would hopefully make the splitting of parameter intervals unnecessary, as approximation can reduce the number of shortcuts. Without splitting, preprocessing time and space are greatly reduced, this comes in handy as they are significantly larger than in the single-criteria scenario.

In general, heuristic CH can help to process graphs that are less hierarchically structured than road networks. In theory, every graph can be preprocessed by CH. But in practice, this is not feasible. The witness searches that decide the necessity of shortcuts take too long, especially when the remaining graph is very dense after a lot of shortcuts have been added. Reducing the number of shortcuts can therefore significantly reduce the preprocessing time.

# 6 Conclusion

We developed an algorithmic idea for an heuristic CH algorithm that guarantees the approximation ratio. Now experiments have to show the feasibility of this approach in practice.